\documentclass[fleqn,twoside]{article}
\usepackage{espcrc2}
\usepackage[figuresright]{rotating}

\newcommand{\rxj}{1RXS\,J170849$-$400910}

\newcommand{\BSAX}{{\em Beppo}SAX}

\input psfig.sty

\title{First evidence of a cyclotron feature in an anomalous X-ray pulsar}
\author{Nanda Rea,\address[UR2]{Physics Department, 
               University of Rome ``Tor Vergata'' \\ 
               Via della ricerca scientifica 1, 00133, Rome, Italy}
                 \address[OAR]{INAF-Osservatorio Astronomico di Roma,\\
               via Frascati 33, 00040, Monteporzio Catone (RM), Italy} 
        Gian Luca Israel,\addressmark[OAR]
        Luigi Stella\addressmark[OAR]
        \thanks{e-mail: nandad@, gianluca@ and stella@mporzio.astro.it}   
        }
\begin{document}

\begin{abstract}
We report on the results of the longest uninterrupted observation of
the Anomalous X-ray Pulsar \rxj\ obtained with the \BSAX\ satellite in
August 2001. The best fit phase-averaged spectrum was found to be an
absorbed power law plus blackbody model, with photon index $\Gamma
\sim 2.4 $ and a black body temperature of $kT_{bb} \sim 0.4
$\,keV. We confirm the presence of significant spectral variations
with the rotational phase of the pulsar. In the spectrum corresponding
to the rising part of the pulse we found an absorption-like feature at
$\sim 8.1$\,keV (significance level of 4$\sigma$), most likely due to
cyclotron resonant scattering. The centroid energy converts to a
magnetic field of $9\times10^{11}$\,G and $1.6\times10^{15}$\,G in the
case of electrons and protons, respectively. If confirmed, this would
be the first detection of a cyclotron feature in the spectrum of an
anomalous X-ray pulsar.

\end{abstract}

\maketitle


\begin{figure}[h]
\psfig{figure=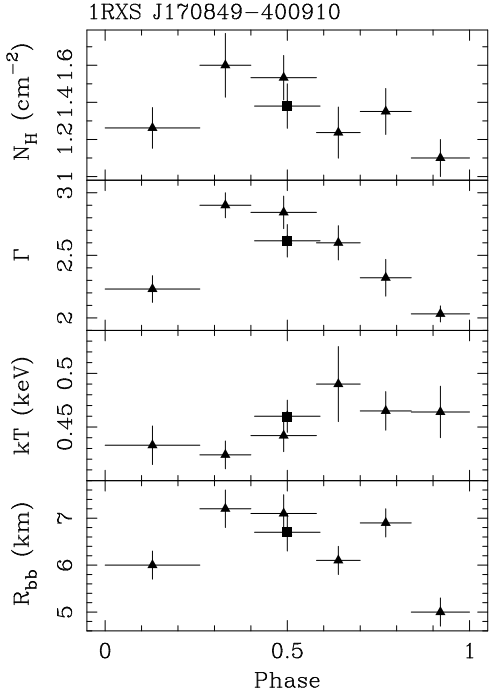,width=13cm,height=10cm,angle=270}
\caption{Spectral parameters variability with phase,
 using an absorbed black body plus power-law model.  Filled squares
 represent the spectral parameters after the addition a of cyclotron line
 model in the 0.4 -- 0.58 phase interval. Note that the relevant points are
 slightly shifted in phase for clarity.}
\end{figure} 


\section{Introduction}

Anomalous X-ray pulsars (AXPs) are characterized by spin periods in
the range of 5-12\,s, steady spin down ($\sim 10^{-11}\,ss^{-1}$),
steep and soft X--ray spectra with luminosities exceding by several
orders of magnitude their spin--down luminosities (Mereghetti \&
Stella 1995).  All five confirmed AXPs lie in the ``galactic'' plane
and two (or three), are associated with supernova remnants. AXPs show
no evidence for a companion and are thus believed to be isolated
neutron stars either having extremely strong magnetic dipole fields
($\sim 10^{14}-10^{15}\,G$; ``magnetars''; Duncan \& Thompson 1992) or
accreting from a residual disk (Alpar 2001).For recent rewiews see
Mereghetti et al. 2002, and references therein.

\rxj\ was discovered with ROSAT (Voges et al. 1996) and $\sim$\,11\,s
pulsations were found in its X--ray flux with ASCA (Sugizaki et
al. 1997). This sourse presents a strange timing behaviour, it is a
quite stable rotator except few glitch type events that occurred in
its secular spin--down (Kaspi et al. 2000, Dall'Osso et al. 2003,
Kaspi \& Gavriil 2003). Searches for an optical counterpart
ruled out the presence of a massive companion (Israel et al. 1999); an
IR counterpart has been recently proposed (Israel et al. 2003). There
is no evidence of pulsed radio emission from the AXP with an upper
limit of 70$\mu$Jy on the pulsation amplitude (Israel et al. 2002).

Here we report on the longest observation (200.0\,ks) of this source,
made with \BSAX\ in August 2001. Analysing spectral data we discovered the
first absorption-like feature ever seen in an AXP, probably due to
cyclotron resonant scattering. For more details about the observation
and the analysis see Rea et al. 2003.



\begin{figure}[htb]
\centerline{\psfig{figure=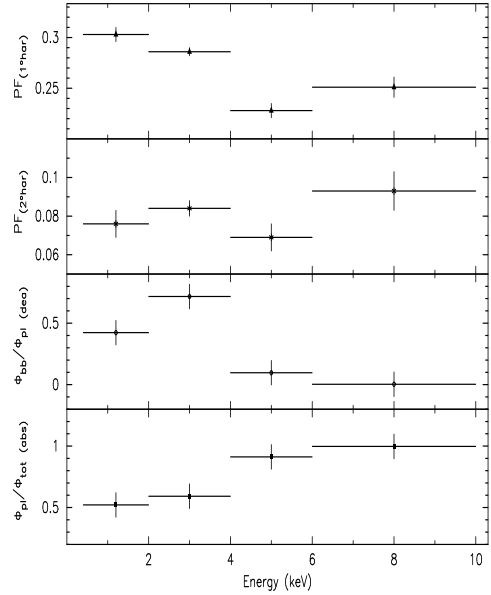,width=8cm,height=9cm,angle=270}}
\caption{The first two panels show the pulsed fractions of the first
and the second harmonics of the spin period in function of energy. In
the third panel is presented the relative intensity of the fluxes of
the power--law and the black body components and in the last one, of
the power--law and the total flux, all in function of energy.}
\end{figure}


\section{Timing and spectral results}

The timing analysis allowed us to estimate a spin period of
$11.000563\pm 0.000005\,s$. The source shows an
energy--dependent profile (Fig.\,3), in particular the pulse minimum
shifts from a phase of $\sim$\,0.0 in the lowest energy light curve
(0.1--2\,keV) to $\sim$\,0.3 in the 6--10\,keV light
curve. Correspondingly, the pulsed fraction decreases from
$\sim$\,30\,\% to $\sim$\,17\,\%.

The energy spectra were well fit with an absorbed blackbody plus a
power law model. The best fit of the phase-average spectrum gave a
reduced $\chi^2$ of 0.95 for 298 degree of freedom (dof) for the
following parameters: column density of $ N_{H} = (1.36 \pm0.06)\times
10^{22}$\,cm$^{-2}$, a blackbody temperature of $kT_{bb}=
0.44\pm0.01$\,keV (blackbody radius of $R_{bb} = 6.6\pm0.4$\,km,
assuming a distance of 5\,kpc) and a photon index of $\Gamma =
2.40\pm0.06$ (all error bars in the text are 90\,\% confidence). The
unabsorbed flux in the 0.5--10\,keV range was $1.87\times
10^{-10}$\,erg\,cm$^{-2}$\,s$^{-1}$ corresponding to a luminosity of
5.6$\times10^{35}$\,erg\,s$^{-1}$ (for a 5 kpc distance). In the
0.5--10 keV band the blackbody component accounts for $\sim$\,30\,\%
of the total unabsorbed flux (see Fig.\,2).

In order to search for spectral features we made a pulse phase
spectroscopy and accumulate spectra in six different phase intervals
(see Fig.\,1 and Tab.\,1). A significant variation of the spectral
parameters with pulse phase was clearly detect (especially for $\Gamma$)
and confirm what previously reported in literature (Israel et
al. 2001). In all intervals but one, an acceptable fit was obtained
with the absorbed power--law plus black body model (reduced $\chi^2$
in the 0.9--1.1 range); a reduced $\chi^2$ of 1.2 was instead obtained
in the 0.4--0.58 phase interval. In the latter case, the data were
systematically below the best fit model in the $\sim$\,7.8--8.4\,keV
range (see Fig.\,4b).


\begin{figure}[h!]
\centerline{\psfig{figure=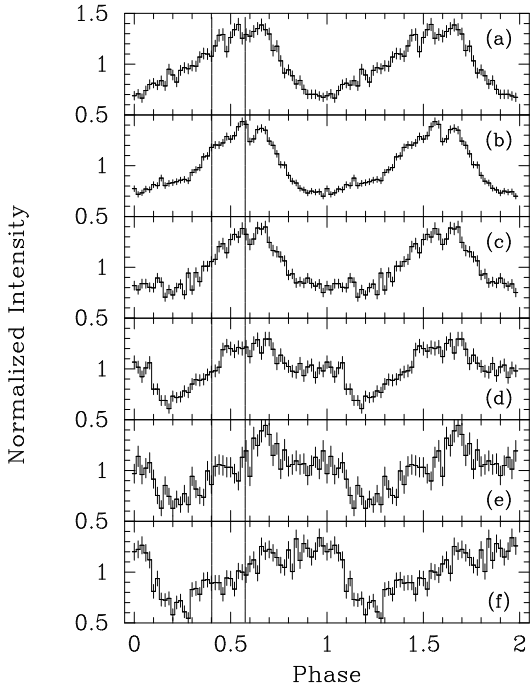,width=10cm,height=14cm} }
\caption{MECS light curves of \rxj\ folded at the best spin period 
(two pulse cycles are shown) 
for six energy bands: (a) 0.1--2 keV; 
(b) 2--3 keV; (c) 3--4 keV; 
(d) 4--5 keV; (e) 5--6 keV; (f) 6--10 keV. The vertical lines mark the 
phase interval in which the absorption-like feature was detected.}
\end{figure}

\newpage

We tried to fit three different models: a Gaussian, an absorption edge
and a cyclotron feature.  While the inclusion of a Gaussian or an
absorption edge did not lead to a significant improvement of the fit,
the cyclotron model (CYCLABS in the XSPEC package, see Mihara et
al. 1990 for details) led to a reduced $\chi^2$ of 1.09, corresponding
to an F-test probability of $1.8\times10^{-5}$ implying to a
single trial significance of 4.5$\sigma$ or 4$\sigma$ after correction
for the six spectra that we analysed (see Fig.\,4 and
Tab\,1).Moreover, taking into account also the possibility of find the
feature in all 1-10 keV range, we found a 3.5$\sigma$ confidence
level.



\section{Discussion}

During the analysis of the longest \BSAX\ observation of \rxj\, we
discovered an absorption-like feature at an energy of $\sim$\,8.1\,keV
in a pulse phase interval corresponding to the rising part of the
$\sim$\,11\,s pulse. This feature was best fit by a resonant cyclotron
feature model with a centroid energy of $\sim$8.1\,keV and an
equivalent width of $\sim$460\,eV (see Fig.\,4 and Tab.\,1).

The detection of an RCF in a specific pulse-phase interval and
superposed to an X-ray continuum that varies with the pulse phase is
reminiscent of the behaviour seen in standard accreting pulsars in
X-ray binaries (Wheaton et al. 1979).  If
interpreted as an electron resonant feature at the base of the
accretion column, the feature at $\sim$\,8.1 keV implies a neutron
star surface magnetic field of $\sim 9.2 \times 10^{11}$\,Gauss (using
a gravitational redshift z=0.3) . This value is just slightly lower
than that measured for electron RCFs in typical accreting X-ray
pulsars (see Fig.\,5); more interestingly it is close to that required
by models for AXPs which involve residual disk accretion in the
spin--down regime. In this context, one can solve the torque equation (see
e.g. Eq. 11.35 in Henrichs 1983) by exploiting the measured value of
$\dot{P}$ and range of accretion luminosity derived from plausible
distances (5--10\,kpc). The surface magnetic (dipole) field obtained
in this way is 0.6--1.1$\times10^{12}$\,G (corrisponding to a fastness
parameter range of $\omega_s = 0.57-0.54$, a typical value for the
spin-down accretion regime; see Ghosh \& Lamb 1979 and Henrichs 1983).

The agreement of this estimate with the magnetic field inferred from
the electron RCF interpretation is intriguing, especially in
consideration of the other analogies with the pulse-phase spectral
dependence of conventional accreting X-ray pulsars. By contrast, if an
electron RCF arose somehow at the polar caps of a rotation powered
pulsar, a B--field strength of 9.2$\times10^{11}$\,G would be in the
range of many radio pulsars and yet much lower than that required to
spin--down at the observed rate through magnetic dipole radiation
($\sim 5 \times 10^{14}$\,G, indeed this was one of the motivations
for magnetar model, see below).

There is a clear correlation between the width and centroid energy of
the electron RCFs in accreting X--ray pulsars (see Fig.\,5, extending
the results of Orlandini \& Dal Fiume 2001). The values from \rxj\ and
SGR\,1806--20 are in good agreement with such a relation. The modest
range of width to centroid energy ratio implied by this indicates that
magnetic field geometry effects at the neutron star surface likely
dominate the RCF width (on the contrary temperature and particle mass
would alter this ratio).  This, in turn, suggests that similar
(relative) ranges of surface magnetic field strength are ``sampled''
by RCFs in accreting X--ray pulsar and, by extension, RCFs in AXPs and
SGRs.

Alternatively the RCF might be due to protons.  For the magnetic field
strengths forseen in the ``magnetar'' scenario, proton cyclotron
features (if any) are expected to lie in the classical X--ray band
(0.1--10\,keV; Zane et al. 2001; Lai \& Ho 2002).  A proton RCF
feature at $\sim 8.1$\,keV would correspond to surface field of
1.6$\times$10$^{15}$\,G (z=0.3). The fact that this value is $\sim 3$
times higher than the surface field derived from the usual magnetic
dipole spin--down formula should not be of concern. According to the
magnetar model, the magnetic field at the star surface and its
vicinity is dominated by higher order multipole field components. At
large radii the dipole component, responsible for the secular
spin--down, dominates.  It is thus expected that a proton RCF feature,
sampling the (total) surface magnetic field strength, provides a
higher value than the mere dipole component.


\begin{figure}[h!]
\centerline{\psfig{figure=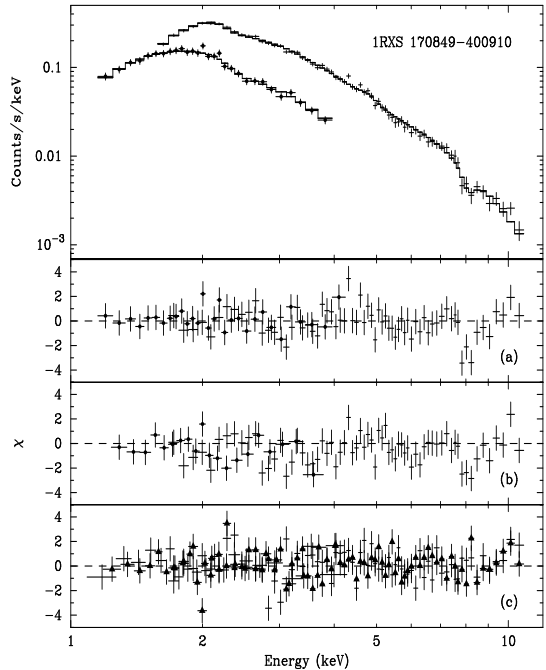,width=8cm,height=13cm} }
\caption{MECS and LECS spectra from the 0.4--0.58 phase interval fit
 with the ``standard model'' (the sum of a blackbody and power law
 with absorption) plus a cyclotron line.  Residuals are relative to
 the ``standard model' alone in order to emphasize the absorption-like
 feature at $\sim 8.1$~keV: (a) the \BSAX\ observations merged
 together; (b) the 2001 observation alone; (c) the phase intervals
 contiguous to that showing the cyclotron absorption feature in the
 merged observations.}
\end{figure}


\begin{table*}
\caption{Spectral parameter values of the phase resolved spectroscopy}
\newcommand{\cc}[1]{\multicolumn{1}{c}{#1}}
\renewcommand{\tabcolsep}{0.3pc} 
\renewcommand{\arraystretch}{1} 
\begin{tabular}{@{}lllllll}
\hline 
Phase & \cc{0.0--0.26} & \cc{0.26--0.4} & \cc{0.4--0.58} & 
\cc{0.58--0.7} & \cc{0.7--0.84} & \cc{$0.84--1.0$} \\
\hline 
$N_{H}(\times10^{22}$\,cm$^{-2})$ & \cc{$1.26\pm0.11$} &
\cc{$1.6\pm0.1$}& \cc{$1.38\pm0.12$} &
\cc{$1.24\pm0.14$} & \cc{$1.35\pm0.12$} & \cc{$1.1\pm0.1$} \\ 
$\Gamma$ & \cc{$2.2\pm0.1$} & \cc{$2.9\pm0.1$} &
\cc{$2.62\pm0.13$}& \cc{$2.60\pm0.13$} & \cc{$2.30\pm0.15$} &
\cc{$2.03\pm0.06$}  \\ 
$kT_{bb}$ (\,keV) & \cc{$0.433\pm0.017$} & \cc{$0.424\pm0.013$} & 
\cc{$0.46\pm0.01$} & \cc{$0.49\pm0.03$} & \cc{$0.465\pm0.018$} & 
\cc{$0.46\pm0.02$}\\ 
$R_{bb}$ (d = 5 kpc; km) & \cc{$ 6.0\pm0.3$ } & \cc{$7.2\pm0.4 $} &
 \cc{$6.7\pm0.4$} & \cc{$6.1\pm0.3$} & \cc{$6.9\pm0.3$} & 
\cc{$5.0\pm0.3$}  \\ 
$E_{cyclabs}$ (\,keV) & \cc{--} & \cc{--} &  \cc{$8.1\pm0.1$} & 
 \cc{--}&  \cc{--} &  \cc{--} \\                           
Line Width (\,keV) & \cc{--} & \cc{--}  &  \cc{$0.2\pm0.1$} &
 \cc{--} &  \cc{--} &  \cc{--} \\ 
Line Depth (\,keV) &  \cc{--} &  \cc{--} & \cc{$0.8\pm0.4$} & 
 \cc{--} & \cc{--} &  \cc{--}\\ 
$\chi^2/d.o.f.$ & \cc{1.08} & \cc{0.98} & \cc{1.09} &  \cc{0.99} & \cc{1.09} & \cc{1.05} \\

\hline
\end{tabular}\\[2pt]
\end{table*}


Other interpretations of the feature at $\sim 8.1$\,keV appear less
likely. Firstly, fitting an edge or line due to photo-electric
absorption provides a less pronounced improvement of the fit than the
RCF model.  Secondly, an edge by iron at a sufficiently large distance
from the neutron star that energy shifts are negligible would require
a high overabundance of this element and intermediate ionisation
stages (such as C--like iron). Yet it has long been known that the
photoionisation equilibrium of such a plasma is unstable (Krolik \&
Kallman 1984; Nagase 1989).  The energy of an ion feature forming in
the neutron star atmosphere would be drastically altered by magnetic
field effects (see Mori \& Hailey 2002 and references therein). In
this and the above interpretations, however, it would also be
difficult to explain why an ion feature is observed only over a
restricted range of pulse phases.

\section{Conclusion}

In conclusion, we found an absorption-like feature in the \BSAX\
X--ray spectrum of \rxj\ taken during the rising phase of the
$\sim$11\,s pulse, which is best fit by a cyclotron resonant
scattering model. If this interpretation is correct, the centroid
energy translates into a model--indipendent magnetic field strength of
$\sim$1.6$\times$10$^{15}$\,G or $\sim$9.2$\times$10$^{11}$\,G
depending on whether protons or electrons, respectively, are
responsible for the feature. 

Further observations and polarization measurements might choose
beetwen the two case. If the line is due to protons this result will
give the confirmation that AXPs are magnetars, if instead, is due to
electrons, this depress the magnetar model in favor of
accretion--driven models for analogy with other accreeting pulsar
magnetic field strenghts known so far.



\newpage 


\begin{figure}[t]
\centerline{\psfig{figure=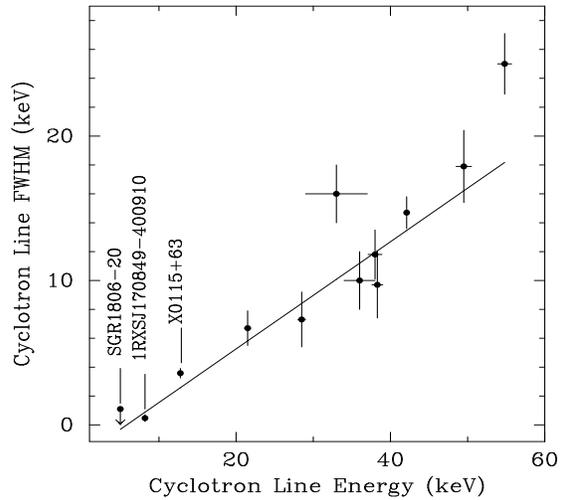,width=10cm,height=12cm} }
\caption{Line width vs. centroid energy from a sample of 
          accreting X-ray pulsars
          with electron RCFs (Orlandini \& Dal Fiume 2001), \rxj\,
         (this Letter) and SGR\,1806--20 (Ibrahim et al. 2002).} 
\end{figure} 


\end{document}